# High-throughput identification of ferromagnetic Kagome candidates in the $AT_6X_4$ and $AT_6X_5$ families

Shiya Chen[1], Vladimir Antropov[2,3†], and Yang Sun[1*]

[1]Department of Physics, Xiamen University, Xiamen 361005, China
[2]Department of Physics and Astronomy, Iowa State University, Ames, IA 50011, USA
[3]Ames National Laboratory, Ames, Iowa 50011, USA

## Abstract

We present a systematic high-throughput density-functional theory study of the thermodynamic stability, collinear magnetic ground states, and electronic structures of layered kagome compounds in the $AT_6X_4$ and $AT_6X_5$ families. Using the experimentally reported structure types as templates, we screened 78 substitutional compositions in each family. Our calculations reproduce the stability and antiferromagnetic character of the known Fe-based Ge compounds and identify six additional stable candidates with robust ferromagnetism. Within collinear spin configurations, we find a clear chemistry-dependent trend: stable Fe-based Ge compounds predominantly adopt AFM2 ground states, whereas stable Mn-based Ge compounds consistently favor ferromagnetic order. Exchange analysis further shows that the magnetic phase space is governed by competing interlayer interactions, consistent with the mechanism established for $AT_6X_6$ kagome magnets. Representative ferromagnetic members from the two structural families also retain kagome-derived dispersive band features near K, although the $AT_6X_5$ phase exhibits stronger band folding and hybridization. Overall, these results establish $AT_6X_4$ and $AT_6X_5$ as promising layered kagome families for realizing ferromagnetism and kagome-derived electronic states.

## I. Introduction

Ferromagnetic (FM) kagome metals have emerged as a paradigmatic platform for exploring the interplay among lattice geometry, magnetism, and electronic structure. In these materials, the combination of time-reversal-symmetry breaking and spin–orbit coupling can generate large Berry curvature, giving rise to remarkable transport phenomena such as anomalous Hall and Nernst effects [1–3]. Representative examples include the massive Dirac fermions observed in $Fe_3Sn_2$ [4] and the large intrinsic anomalous Hall response associated with Weyl points in $Co_3Sn_2S_2$ [5]. Beyond these notable phenomena, a central materials challenge is to understand how chemical composition controls thermodynamic stability and magnetic order, which is essential for identifying robust FM kagome compounds.

†Email: antropov@ameslab.gov
*Email: yangsun@xmu.edu.cn

Among layered kagome intermetallics, the $AT_6X_6$ family (A = spacer cation, T = transition metal, X = p-block element) has been extensively investigated and is now known to host a wide variety of magnetic states [6–9]. In comparison, two structurally related but much less explored families, $AT_6X_4$ ($LiFe_6Ge_4$-type) and $AT_6X_5$ ($LiFe_6Ge_5$-type), were first reported by Welk and Schuster [10]. These compounds retain the same kagome-layer backbone as $AT_6X_6$ but adopt distinct interlayer stacking geometries that can substantially modify the magnetic interactions. Notably, the transition-metal sublattice in $AT_6X_4$ is crystallographically equivalent to the bilayer kagome framework of $Fe_3Sn_2$ [11], making the $AT_6X_4$ and $AT_6X_5$ families natural extensions of the layered kagome design space with tunable interlayer coupling.

Experimentally, known members of these families remain concentrated in the Fe–Ge series: $LiFe_6Ge_4$, $ScFe_6Ge_4$, $ZrFe_6Ge_4$ [10,12], and $LiFe_6Ge_5$ [10]. Magnetic measurements have confirmed antiferromagnetic (AFM) ordering in $LiFe_6Ge_4$ and $LiFe_6Ge_5$ [13], while $ScFe_6Ge_4$ exhibits a high Néel temperature of approximately 650 K [11]. First-principles calculations further established that interlayer exchange competition governs the collinear AFM ground states of $LiFe_6Ge_4$ and $LiFe_6Ge_5$ [14]. Although Vishina et al. [15] theoretically demonstrated that substituting Ge with Al or Ga in $LiFe_6Ge_4$-derived compounds can stabilize an FM ground state with promising magnetization and magnetocrystalline anisotropy, no FM member of the $AT_6X_4$ or $AT_6X_5$ families has yet been experimentally confirmed. This contrasts sharply with the $AT_6X_6$ family, where FM kagome metals such as $TbMn_6Sn_6$ [16,17] and $LiMn_6Sn_6$ [3,18,19] are well established. The absence of FM order in $AT_6X_4$ and $AT_6X_5$, despite their close structural relationship to FM-prone Mn-based kagome systems, raises a fundamental question: can chemical substitution, especially replacing Fe with Mn, induce ferromagnetism in these layered kagome frameworks?

In this work, we conduct a systematic high-throughput DFT investigation of the thermodynamic and magnetic stability of the $AT_6X_4$ and $AT_6X_5$ kagome families. Motivated in part by our previous high-throughput study of the related $AT_6X_6$ kagome compounds, which established clear chemistry-dependent stability and magnetic trends in the $AT_6X_6$ family, we extend a similar screening strategy to two less-explored structural families based on experimentally reported prototypes. This survey not only reproduces known tendencies in Fe-based compounds, but also identifies several previously unreported stable FM candidates, thereby providing a practical roadmap for future experimental exploration. Focusing on representative predictions, we further examine the exchange-interaction picture underlying their magnetic behavior and assess whether their electronic structures retain characteristic kagome-derived band features.

## II. METHODS

The substituted $AT_6X_4$ and $AT_6X_5$ structures were optimized by spin-polarized density functional theory (DFT) calculations using the VASP code [20,21], which employs the projector augmented wave (PAW) method [22]. The exchange and correlation energy is treated with the spin-polarized generalized gradient approximation (GGA) and parameterized by the Perdew−Burke−Ernzerhof formula (PBE) [23]. A plane-wave basis was used with a kinetic energy cutoff of 520 eV, and the convergence criterion for the total energy was set to $10^{-5}$ eV. Monkhorst−Pack's sampling scheme [24] was adopted for Brillouin-zone sampling with a k-point spacing of $2\pi \times 0.033$ Å$^{-1}$. The lattice and atomic coordinates are fully relaxed until the force on each atom is less than 0.01 eV/Å. All geometry optimizations were initialized in the FM configuration. The converged energies were used for the formation-energy and convex-hull analyses regardless of the final moment alignment. The FM–AFM energy differences were obtained by comparing the self-consistent energies of the FM, AFM1, and AFM2 configurations on the same PBE-relaxed structure for each compound. FM-initialized calculations that did not retain a uniform same-sign FM alignment were excluded from the FM–AFM energy comparison. The energies of the different magnetic states were obtained from self-consistent calculations performed directly on the PBE-relaxed structures. This common high-throughput approach assumes that the magneto-volume effect is negligible for the primary screening of the magnetic ground state. SOC was included only in the post-processing of electronic structures.

We characterized the thermodynamic stability of each $AT_6X_4$ and $AT_6X_5$ compound by calculating its energy above the convex hull $E_{\mathrm{d}}$, which represents the energy difference against the most stable competing reference phases. The competing reference phases for the convex-hull construction were identified from the Materials Project database [25]. Their total energies were then recalculated using the same PBE-based computational setup employed in this work, ensuring an internally consistent dataset for the formation-energy and energy-above-hull analysis.

The Heisenberg model stability parameters have been obtained using the code from Refs. [26,27], with the OpenMX package. For the OpenMX calculations, the SCF step used a k-point spacing of $2\pi \times 0.025$ Å$^{-1}$, while the exchange calculations employed a denser spacing of $2\pi \times 0.01$ Å$^{-1}$, which was verified to be converged.

## III. Results and discussion

### 1. Stability analysis

We adopt the experimentally reported $LiFe_6Ge_4$- and $LiFe_6Ge_5$-type structures as structural templates for systematic elemental substitutions in the $AT_6X_4$ and $AT_6X_5$ families. Both structure types share a layered kagome framework, in which transition-metal atoms form two-dimensional kagome networks separated by spacer layers composed of A and X atoms. In the $AT_6X_4$ structure (Fig .1(a)(b)), the kagome layers are stacked along the c axis as characteristic bilayer units, with adjacent bilayers separated by AX2 spacer layers. The $AT_6X_5$ structure can be viewed as a derivative of the $AT_6X_4$ structure (Fig.1 (c)(d)), distinguished by the insertion of an additional X2 layer into the spacer block.

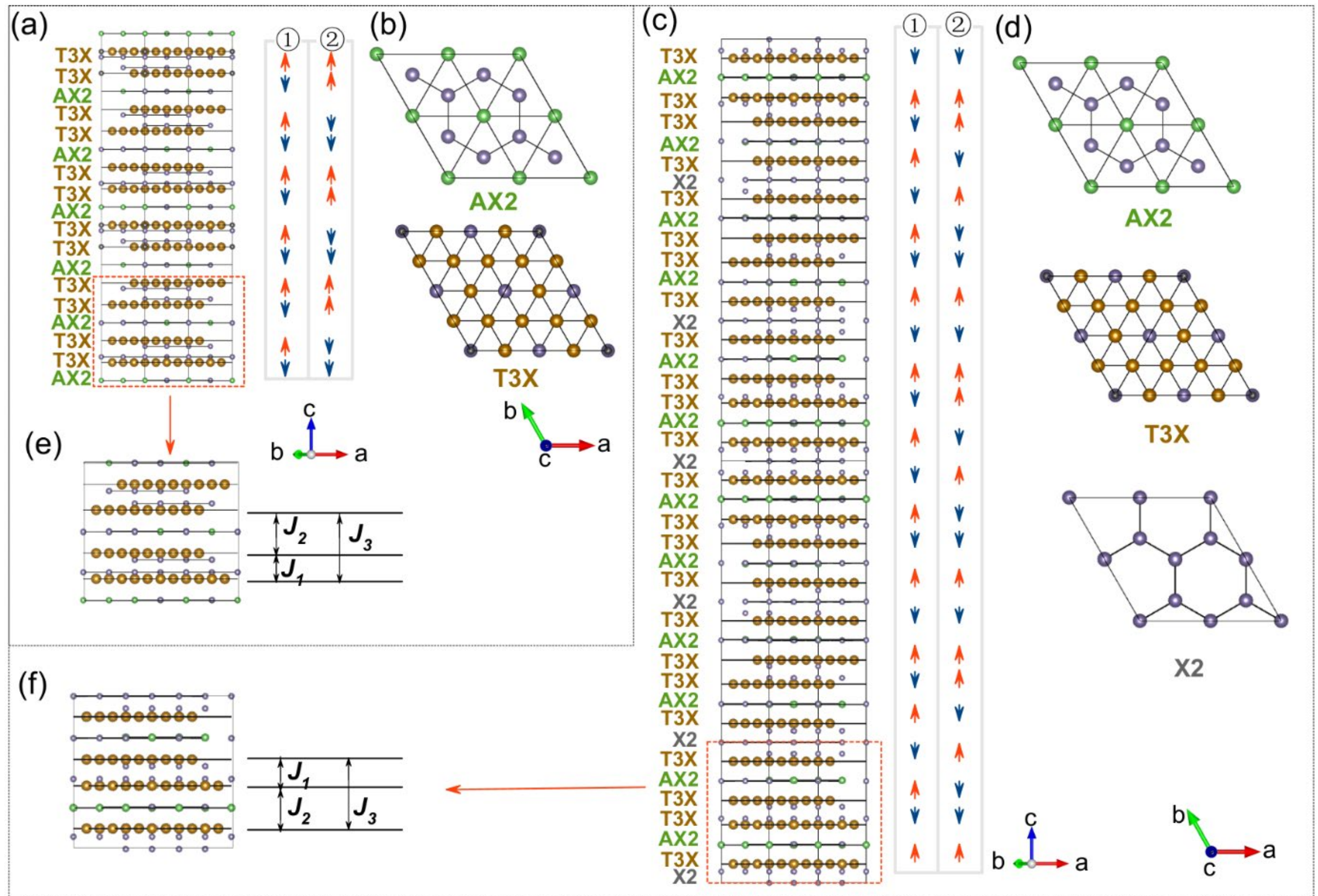

FIG. 1 Crystal structures, representative collinear interlayer magnetic stackings, and definitions of interlayer interaction parameters for the layered kagome compounds $AT_6X_4$ and $AT_6X_5$. (a) Side view of the crystal structure of $AT_6X_4$, showing the stacking of T3X and AX2 layers along the stacking direction. The two spin sequences labeled 1 and 2 denote the two representative collinear interlayer antiferromagnetic configurations, AFM1 and AFM2, respectively. (c) Side

view of the crystal structure of $AT_6X_5$ together with the corresponding AFM1 and AFM2 stacking patterns; compared with $AT_6X_4$, an additional X2 layer is present. (b,d) Top views of the constituent structural motifs: AX2 and T3X for $AT_6X_4$, and AX2, T3X, and X2 for $AT_6X_5$. (e,f) Definition of the interlayer T–T interaction parameters $J_1$, $J_2$,, and $J_3$, which denote the first-, second-, and third-nearest interlayer interactions, respectively, for $AT_6X_4$ and $AT_6X_5$. Different collinear magnetic stackings can be distinguished by the relative spin alignment between neighboring kagome layers associated with these $J_n$. In all cases, spins are ferromagnetically aligned within each kagome layer. A, T, and X atoms are shown in green, brown, and purple, respectively.

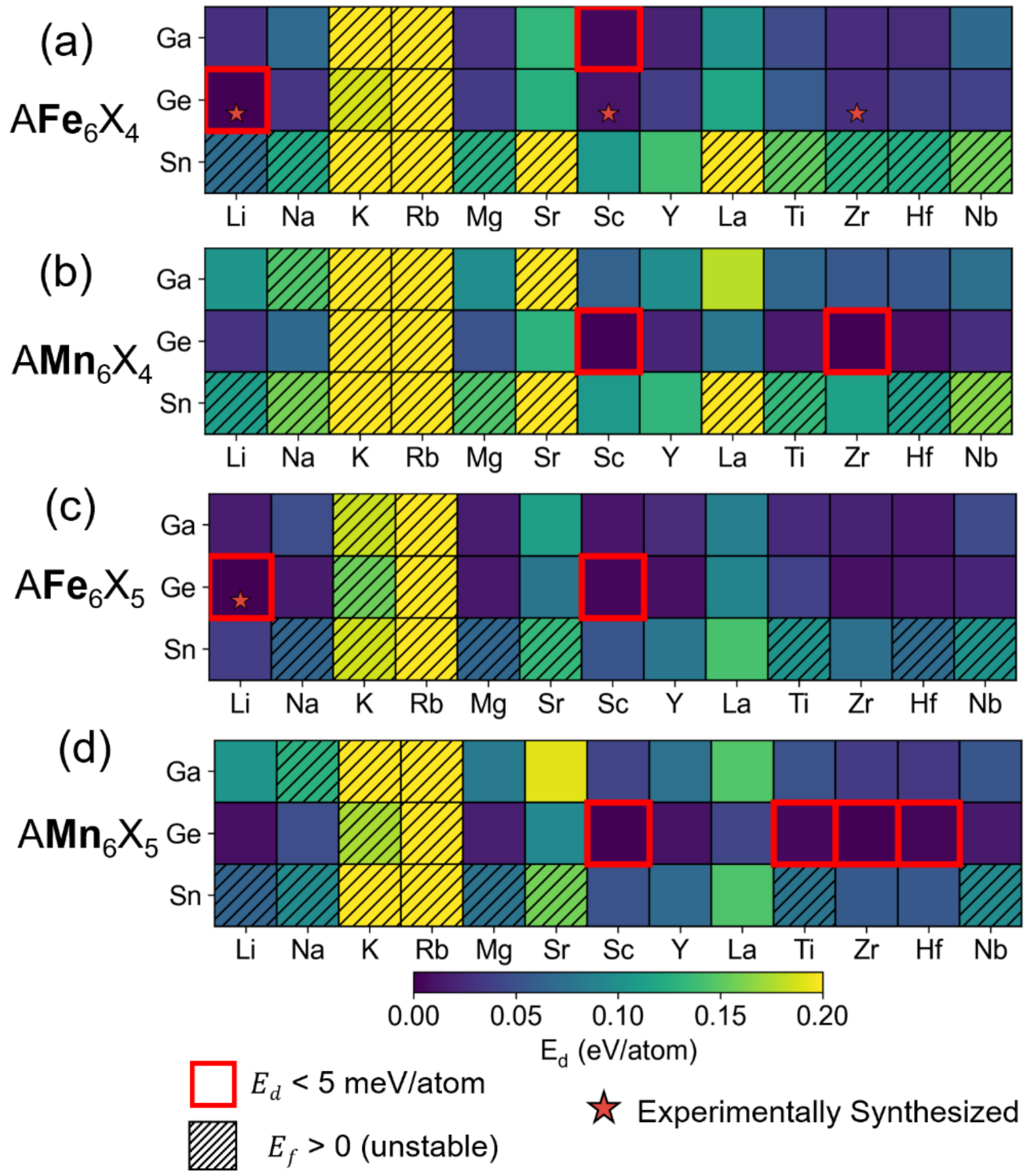

FIG. 2. Thermodynamic stability maps for $AT_6X_4$ and $AT_6X_5$ kagome compounds with T = Fe or Mn. (a–d) Color scale shows the energy above hull $E_d$ (eV/atom) for (a) $AFe_6X_4$, (b) $AMn_6X_4$, (c) $AFe_6X_5$, and (d) $AMn_6X_5$. Compounds with $E_d < 5$ meV/atom are considered stable and are marked by red boxes; red stars indicate experimentally synthesized compounds. The horizontal axis lists A-site elements from Li to Nb, and the vertical axis lists X-site elements in the order Ga, Ge, and Sn.

For the compositional screening, we adopt the same elemental set as in our previous work, restricting the transition-metal site T to Fe and Mn, the X site to Ga, Ge, and Sn, and the A site to elements from Li to Nb. This combinatorial design yields 78 compositions for each structural prototype. The stability results for all systems are summarized in Fig. 2. Following our previous work and earlier high-throughput studies [28,29], we classify compounds with $5 < E_d < 100$ meV/atom as metastable and those with $E_d < 5$ meV/atom as stable. In the following, we first discuss the broader metastable landscape and then focus on the much smaller subset of strictly stable compounds.

As shown in Fig. 2, the thermodynamic stability of the $AT_6X_4$ and $AT_6X_5$ families is strongly influenced by both A-site and X-site chemistry, without a simple overall tendency for Fe-based systems to be systematically more or less stable than Mn-based ones. Using $E_d < 100$ meV/atom as the metastability criterion, we identify a broad pool of candidate compounds across both structural families. A complete list of these compounds is provided in Table S1.

Beyond the compounds satisfying the strict $E_d < 5$ meV/atom criterion, 18 systems lie within 5–20 meV/atom above the convex hull, including 17 predicted candidates and the experimentally synthesized $ScFe_6Ge_4$. This low-energy range is experimentally relevant: a large-scale survey of 29,902 experimentally reported inorganic crystalline phases found a median metastability of approximately 15 meV/atom and a 90th-percentile value of approximately 67 meV/atom, demonstrating that experimentally realized compounds often lie above the 0 K convex hull [30]. A representative benchmark discussed in Ref. [31] reported an uncertainty of approximately 24 meV/atom for DFT relative polymorph energies. Accordingly, hull energies of 5–20 meV/atom do not necessarily preclude experimental realization and should be regarded as low-energy metastable candidates. Nevertheless, proximity to the hull does not guarantee synthesizability, which also depends on kinetic accessibility, finite-temperature effects, vibrational stability, phase competition, and synthesis conditions.

The experimentally reported members are currently concentrated in the Fe-based Ge series, including $LiFe_6Ge_4$, $ScFe_6Ge_4$, $ZrFe_6Ge_4$, and $LiFe_6Ge_5$, all of which our calculations reproduce as stable or metastable, consistent with experiment. The Li–Fe–Ge series illustrates that low hull energy does not necessarily imply straightforward single-phase synthesis. Using the same LiH-based route, $LiFe_6Ge_4$ and $LiFe_6Ge_5$ were obtained with phase fractions of 92.9 and 91.0 wt.%, respectively, with each phase appearing as the main impurity in the other sample, whereas $LiFe_6Ge_6$ was obtained as a single-phase product [13]. This mutual phase coexistence highlights the importance of precise control over stoichiometry and synthesis conditions for the $AT_6X_4$/$AT_6X_5$ phases.

Turning to the predicted stability landscape, a more selective picture emerges when attention is restricted to the stable compounds with $E_d < 5$ meV/atom. Within this regime, several Mn-based Ge members stand out as the most favorable, with six new stable systems predicted: $ScMn_6Ge_4$, $ZrMn_6Ge_4$, $ScMn_6Ge_5$, $TiMn_6Ge_5$, $ZrMn_6Ge_5$, and $HfMn_6Ge_5$. This suggests that Mn-containing $AT_6X_4$ and $AT_6X_5$ compounds are especially promising synthetic targets. $ScFe_6Ga_4$ is the only stable Fe-based Ga compound in our dataset and thus stands out from the predominantly Mn-based Ge FM candidates identified here. Nearly all Sn-based compositions lie outside the stable window, unlike the $AT_6X_6$ family studied previously, where Sn-based systems were found to be nearly as stable as the Ge- and Ga-based ones. In addition, K- and Rb-containing compounds are consistently located in the unstable region in all four panels, indicating that these alkali-metal compositions are generally unfavorable for stabilizing the layered kagome phases.

Comparison with the corresponding $AT_6X_6$ results reveals a systematic stoichiometric trend (see Fig. S1). Across the 78 composition-matched sets, $AT_6X_6$ is the most stable in every case, and 76 follow the full stability ordering $E_d(AT_6X_4) \geq E_d(AT_6X_5) \geq E_d(AT_6X_6)$, indicating that thermodynamic stability generally decreases with progressive X removal. For every A element considered, the $AFe_6Ge_n$ (n = 4–6) series follows the same ordering, with $AFe_6Ge_5$ intermediate in stability between $AFe_6Ge_6$ and $AFe_6Ge_4$. Because changing the stoichiometry simultaneously modifies both electron count and layer stacking, their individual contributions cannot be separated within the present calculations; nevertheless, electron-count tuning and alternative stacking arrangements may offer routes to stabilize $AFe_6Ge_5$.

The Fe–Mn stability difference is likewise X-dependent rather than universal. In the $AT_6X_4$ and $AT_6X_5$ families, Fe is consistently favored in Ga-based pairs, Ge-based pairs are nearly balanced, and Sn-based pairs show only a weak Fe preference. Thus, Mn substitution does not generally improve thermodynamic stability; its main advantage emerges in the stable Ge subset, where it is associated with robust FM order rather than the AFM2 order favored by the Fe counterparts, as discussed below.

In our previous study of the structurally related $AT_6X_6$ family, phonon and elastic-constant calculations further confirmed the dynamical and mechanical stability of the representative stable ferromagnets $LiMn_6Ge_6$ and $NaMn_6Ge_6$ [32]. These results support the general viability of this layered kagome chemistry and provide additional motivation for experimental investigation of the present $AT_6X_4$ and $AT_6X_5$ candidates.

## 2. Collinear Magnetic Ground States

Having established the thermodynamic landscape, we next focus on the collinear magnetic ground states of the experimentally relevant stable and low-energy metastable compositions. To determine the collinear magnetic ground state, we compared the energies of an FM state against several AFM configurations. These configurations, denoted AFM1 and AFM2, are defined by their interlayer stacking sequence along the c-axis. All of them retain FM alignment within the kagome planes, as schematically shown in the right panels of Fig. 1(a) and 1(c).

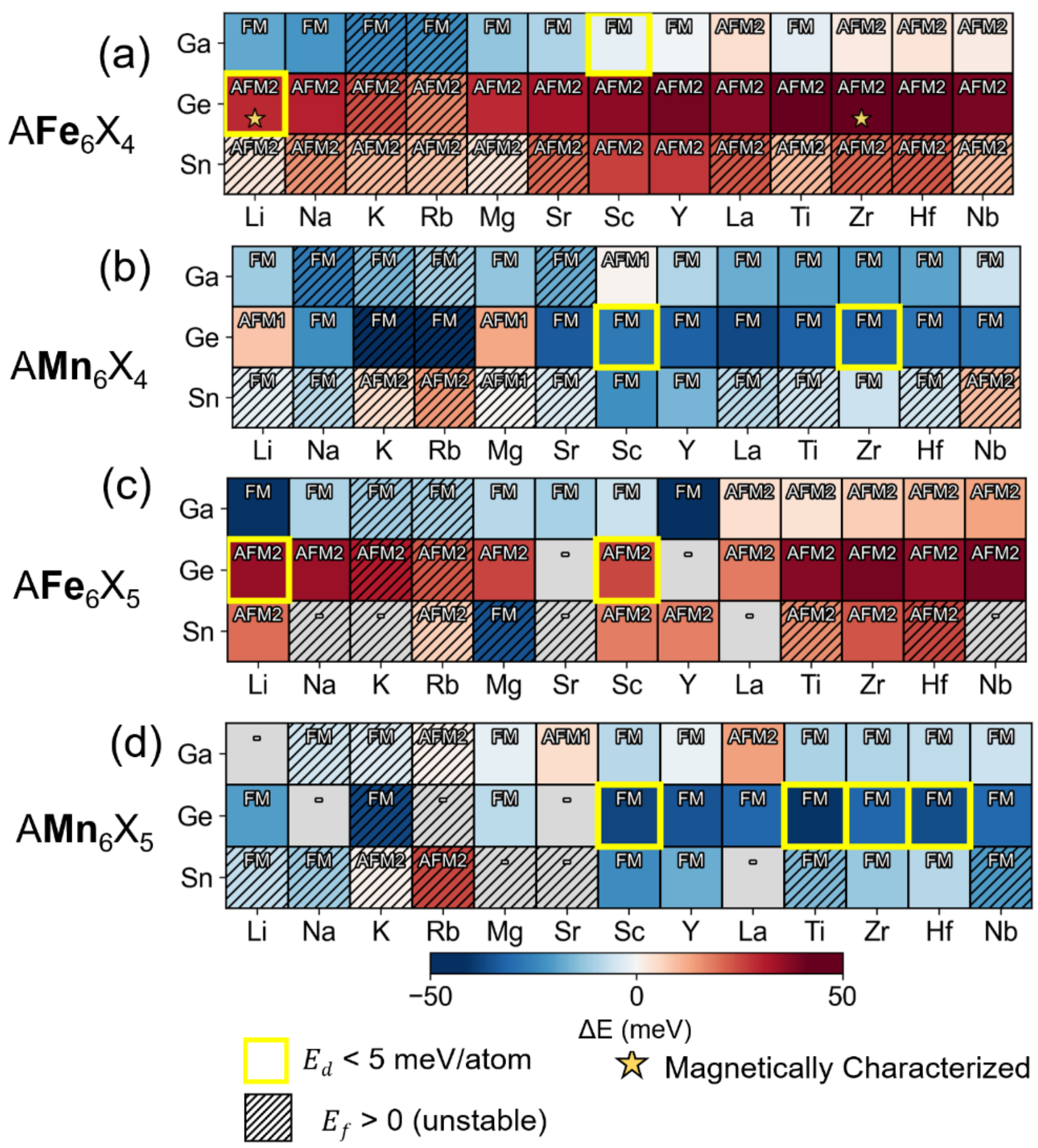

FIG. 3. Magnetic ground states and FM–AFM energy competition for the same $AT_6X_4$ and $AT_6X_5$ families. (a–d) Color scale shows the magnetic energy difference between the FM state and the lowest-energy AFM state, $\Delta E = E(FM) - E_{min}(AFM)$, in meV per transition-metal atom (negative favors FM; positive favors AFM), for (a) $AFe_6X_4$, (b) $AMn_6X_4$, (c) $AFe_6X_5$, and (d) $AMn_6X_5$. The label in each cell denotes the calculated magnetic ground state (FM, AFM1, or AFM2). Compounds with $E_d < 5$ meV/atom are highlighted by yellow boxes, and yellow stars mark magnetically characterized compounds. The horizontal axis lists A-site elements from Li to Nb, and the vertical axis lists X-site elements in the order Ga, Ge, and Sn. Gray cells indicate systems excluded from magnetic comparison because the FM-initialized calculations did not converge to FM states.

Among the 78 $AT_6X_5$ systems, 14 FM-initialized calculations converged to mixed-sign, nearly compensated configurations instead of retaining a uniform FM alignment, whereas all 78 $AT_6X_4$ calculations preserved the same-sign FM alignment. These 14 $AT_6X_5$ systems are shown in gray in Fig. 3 and were excluded from the FM–AFM energy comparison.

As shown in Fig. 3, the calculated collinear magnetic ground states of the $AT_6X_4$ and $AT_6X_5$ families exhibit clear chemistry-dependent trends. Fe-based Ge compounds consistently favor AFM2 order, including the experimentally known members $LiFe_6Ge_4$, $ScFe_6Ge_4$, $ZrFe_6Ge_4$, and $LiFe_6Ge_5$. Table 1 shows that all four adopt AFM2 ground states with positive $\Delta E$ values of 29.5, 36.4, 51.1, and 35.4 meV, respectively, indicating robust collinear antiferromagnetism. These results are also consistent with available experimental reports, which have established the AFM nature of $LiFe_6Ge_4$, $ScFe_6Ge_4$, and $LiFe_6Ge_5$ [11,13]. Although the detailed magnetic structures of $LiFe_6Ge_4$ and $LiFe_6Ge_5$ have not yet been experimentally resolved, our AFM2 assignment is consistent with previous theoretical predictions [14]. In contrast, the only stable Ga-based compound, $ScFe_6Ga_4$, appears to be magnetically much less robust. Its very small $\Delta E$ of -2 meV indicates a near-degenerate FM–AFM competition rather than a strongly stabilized FM ground state.

A markedly different trend is found in the Mn-based systems. In both the $AT_6X_4$ and $AT_6X_5$ families, all stable Mn-based Ge compounds highlighted in Fig. 3 favor FM ground states. Specifically, $ScMn_6Ge_4$, $ZrMn_6Ge_4$, $ScMn_6Ge_5$, $TiMn_6Ge_5$, $ZrMn_6Ge_5$, and $HfMn_6Ge_5$ are predicted to be both stable and FM, with sizable negative $\Delta E$ values ranging from −27.7 to −40.8 meV. This behavior stands in clear contrast to the Fe-based Ge members, which preferentially stabilize AFM2 order, and indicates that the Mn-based Ge systems provide a favorable platform for robust collinear FM.

More generally, as discussed above, Sn-based members in both structural families are thermodynamically unstable and therefore less relevant as realistic candidates, even though some still exhibit well-defined collinear magnetic solutions in our calculations. In addition, AFM1 appears only in a few scattered cases and is rarely associated with thermodynamically stable compounds. Overall, combining the thermodynamic and magnetic screening results reveals a clear contrast between the Fe- and Mn-based systems: stable Fe-based Ge compounds mainly favor robust AFM2 order, whereas stable Mn-based Ge compounds consistently favor robust FM order.

We further compared composition-matched $AT_6X_6$, $AT_6X_5$, and $AT_6X_4$ systems that passed the final-moment quality control (Figs. S2 and S3). The FM–AFM energy difference does not exhibit a regular monotonic evolution across the three structural families, indicating that the magnetic stability is not simply controlled by the structural family. Nevertheless, the local FM moments show a modest overall increase from $AT_6X_6$ to $AT_6X_5$ and $AT_6X_4$, although the differences between $AT_6X_5$ and $AT_6X_4$ remain relatively small and chemistry dependent.

**Table 1.** Stabilities and magnetic properties of experimentally synthesized (marked with *) and newly predicted stable ($E_d < 5$ meV/atom) $AT_6X_4$ and $AT_6X_5$ compounds. The table lists magnetic moments and the magnetic ground state, with experimentally reported states shown in parentheses. Magnetic moments are given in $\mu_B$ per transition metal atom. The energy difference is calculated as $\Delta E = E(FM) - E_{min}(AFM)$, where $E_{min}(AFM)$ is the energy of the lowest-energy AFM configuration (AFM if $\Delta E > 0$ and FM if $\Delta E < 0$.)

| No. | System | $E_d$ (meV/atom) | Magnetic Ground State | Moment ($\mu_B$) | $\Delta E$ (meV) |
|---|---|---|---|---|---|
| 1 | $ScFe_6Ge_4$ | 10.6* | AFM2 (AFM2) | 2.100 | 36.4 |
| 2 | $ZrFe_6Ge_4$ | 26.3* | AFM2 | 2.181 | 51.1 |
| 3 | $LiFe_6Ge_4$ | 0* | AFM2 (AFM) | 2.031 | 29.5 |
| 4 | $LiFe_6Ge_5$ | 0* | AFM2 (AFM) | 1.830 | 35.4 |
| 5 | $ScFe_6Ga_4$ | 4.7 | FM | 2.171 | -2 |
| 6 | $ScFe_6Ge_5$ | 2.7 | AFM2 | 2.024 | 25.2 |
| 7 | $TiMn_6Ge_5$ | 4.5 | FM | 2.111 | -40.8 |
| 8 | $HfMn_6Ge_5$ | 2.9 | FM | 2.132 | -36.3 |
| 9 | $ScMn_6Ge_4$ | 1.2 | FM | 2.130 | -27.7 |
| 10 | $ScMn_6Ge_5$ | 0 | FM | 2.046 | -37.6 |
| 11 | $ZrMn_6Ge_4$ | 0 | FM | 2.174 | -32.1 |
| 12 | $ZrMn_6Ge_5$ | 0 | FM | 2.141 | -31.6 |

## 3. Exchange coupling analysis

Table 2. Interlayer T–T stability parameters $J_n$ (n=1–3) for representative stable kagome compounds $ScFe_6Ga_4$ (weak ferromagnetism), $ScMn_6Ge_4$ (robust ferromagnetism), and $ScFe_6Ge_5$ (robust antiferromagnetism). The magnetic ground state and the corresponding local magnetic moments, calculated using the GGA and LDA functionals, are listed in units of $\mu_B$ per transition-metal atom. Here, a positive $J_n$ indicates that the corresponding interlayer alignment is stable, whereas a negative $J_n$ indicates an unstable configuration. The equivalent Heisenberg exchange parameters can be obtained by applying a minus sign for antiparallel pair alignment and normalizing by $S_i S_j = M_i M_j / 4$.

| System | Order | $J_1$ (meV) | $J_2$ (meV) | $J_3$ (meV) | Moment GGA ($\mu_B$) | Moment LDA ($\mu_B$) | $\Delta E$ (meV) |
|---|---|---|---|---|---|---|---|
| $ScFe_6Ga_4$ | FM | 188 | -28 | -9 | 2.171 | 1.979 | -2 |
| $ScMn_6Ge_4$ | FM | 78 | 47 | 11 | 2.130 | 1.954 | -27.7 |
| $ScFe_6Ge_5$ | AFM2 | 29 | 35 | 4 | 2.024 | 1.647 | 25.2 |

To further understand the microscopic origin of the magnetic trends shown in Fig. 3, we analyze three representative stable compounds selected to capture the three distinct regimes identified in our screening: $ScFe_6Ga_4$ as a weak FM representative, $ScMn_6Ge_4$ as a robust FM representative, and $ScFe_6Ge_5$ as a robust AFM representative. This analysis follows the same rationale as in our previous $AT_6X_6$ study [32], where the competition among interlayer exchange interactions was shown to govern the stability of different collinear magnetic states. The calculated local moments remain close to 2 $\mu_B$ per transition-metal atom in both GGA and LDA, indicating that the differences in magnetic ground states arise primarily from the hierarchy of interlayer exchange interactions rather than from substantial changes in moment magnitude.

Among all newly predicted stable compounds in the present $AT_6X_4$ and $AT_6X_5$ families, $ScFe_6Ga_4$ is particularly noteworthy because it is the only system that exhibits a very small magnetic energy difference ($\Delta E = -2$ meV). Its interlayer exchange hierarchy is strongly competing, with a large positive $J_1$ but negative $J_2$ and $J_3$, indicating that the FM state is only weakly stabilized by the leading interlayer interactions. These features identify $ScFe_6Ga_4$ as the most promising stable candidate in our dataset for future noncollinear verification, because its very small FM–AFM energy difference and competing interlayer couplings place it closest to a possible noncollinear regime within the present screening.

By contrast, $ScMn_6Ge_4$ exhibits a robust FM ground state, with a sizable $\Delta E = -27.7$ meV and uniformly positive $J_1$, $J_2$, and $J_3$, showing that all major interlayer couplings cooperatively support the same FM alignment. A similarly consistent AFM pattern is found in $ScFe_6Ge_5$ whose AFM2 ground state is associated with $\Delta E = 25.2$ meV and positive $J_n$ values throughout, indicating that the realized AFM2 stacking is microscopically well supported by dominant interlayer interactions.

When compared with our previous $AT_6X_6$ study [32], these results suggest that the present $AT_6X_4$ and $AT_6X_5$ systems are governed by a closely related magnetic mechanism. In this picture, a small $\Delta E$ reflects strong competition among different interlayer exchange channels and signals proximity to magnetic frustration or even noncollinear order, whereas a large $\Delta E$ indicates a coherent exchange hierarchy in which the dominant interactions consistently stabilize a well-defined collinear ground state. Therefore, $ScFe_6Ga_4$ represents a competition-dominated regime close to noncollinearity, while $ScMn_6Ge_4$ and $ScFe_6Ge_5$ correspond to robust collinear FM and AFM regimes, respectively.

## 4. Electronic structure analysis

To probe the electronic consequences of the two layered-kagome structural prototypes, we selected $ScMn_6Ge_4$ and $ScMn_6Ge_5$ as a chemically matched pair of representative FM compounds from the $AT_6X_4$ and $AT_6X_5$ families, respectively. Both systems are thermodynamically stable and exhibit robust FM ground states, while keeping the A, T, and X chemistry closely matched. This choice allows us to compare the kagome-derived band features of the two structural types while minimizing changes arising from chemical substitution. As shown in Fig. 4, both compounds display spin-polarized metallic band structures with several dispersive crossings near the K point. In $ScMn_6Ge_4$, a clear Dirac point (DP) and an additional higher-energy DP-like feature can be identified near K, while a relatively weakly dispersive kagome-derived band appears around −0.4 eV. Notably, the lower-energy DP in $ScMn_6Ge_4$ remains essentially gapless after accounting for SOC, making it an especially interesting kagome-band candidate within the present dataset.

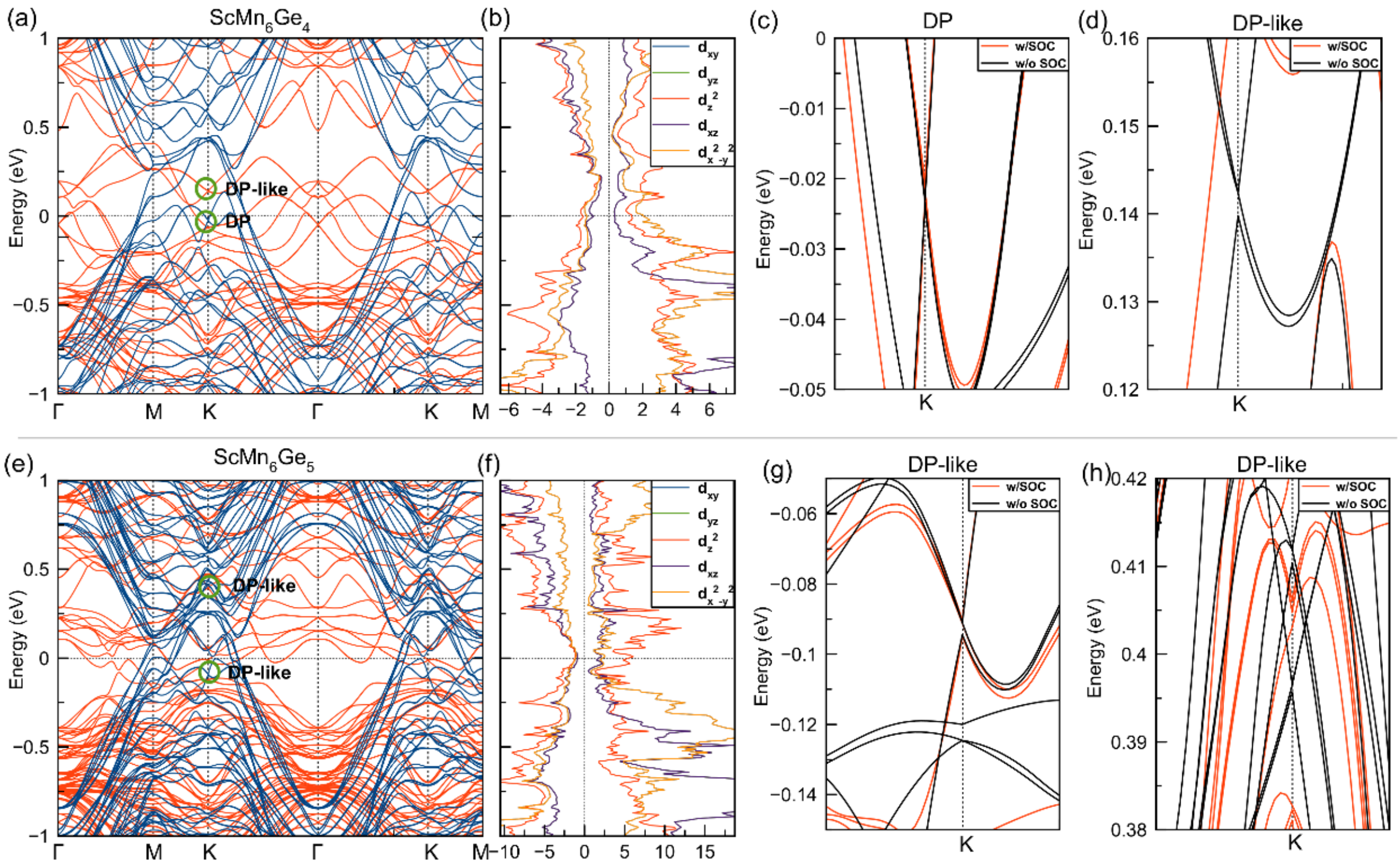

FIG. 4. Spin-polarized electronic structures of FM $ScMn_6Ge_4$ (top row) and $ScMn_6Ge_5$ (bottom row). (a,e) Spin-resolved band structures calculated without SOC, where red (blue) bands denote the majority (minority) spin channel; representative dispersive crossings and Dirac-point/Dirac-point-like (DP/DP-like) features are highlighted by green circles. (b,f) Mn-3d orbital–projected density of states (DOS) for $ScMn_6Ge_4$ (b) and $ScMn_6Ge_5$ (f). (c,d) Enlarged views of the DP and DP-like features near the K point for $ScMn_6Ge_4$, comparing calculations with SOC (red) and without SOC (black). (g,h) Corresponding SOC/no-SOC zoom-ins near K for $ScMn_6Ge_5$ (DP-like features).

Compared with $ScMn_6Ge_4$, the more complex layer stacking in $ScMn_6Ge_5$ leads to stronger band folding and interband hybridization. Nevertheless, DP-like features can still be resolved around K, indicating that key kagome-derived dispersions are retained. Unlike the sharper Dirac crossing in $ScMn_6Ge_4$, however, the K-point features in $ScMn_6Ge_5$ are better described as DP-like rather than ideal Dirac points, reflecting the stronger hybridization induced by the more complex $AT_6X_5$ stacking. The orbital-resolved DOS further reveals a clear difference in orbital character between the two systems. In $ScMn_6Ge_5$, the Mn $d_{x^2-y^2}$ and $d_{z^2}$ orbitals contribute at comparable levels near the Fermi energy, whereas in $ScMn_6Ge_4$ the $d_{z^2}$ contribution is more pronounced. Overall, both $ScMn_6Ge_4$ and $ScMn_6Ge_5$ retain representative kagome-derived dispersive features near K, with $ScMn_6Ge_4$ emerging as the cleaner electronic candidate because of its sharper low-energy Dirac-like feature and weaker SOC-induced modification. However, compared with the $AT_6X_6$ ferromagnets $LiMn_6Ge_6$ and $NaMn_6Ge_6$ discussed in our previous work [32], we do not identify a well-isolated Dirac–VHS–flat-band manifold or sizable SOC-gapped Dirac cones close to the Fermi level in the present $AT_6X_4$/$AT_6X_5$ compounds. Therefore, their topological features near the Fermi level are less pronounced than those in the $AT_6X_6$ systems.

## IV. Conclusions

In summary, our high-throughput DFT study reveals coupled thermodynamic and magnetic trends across the layered $AT_6X_6$, $AT_6X_5$, and $AT_6X_4$ kagome families. Thermodynamic stability generally decreases from $AT_6X_6$ to $AT_6X_5$ and $AT_6X_4$, indicating that progressive X removal is unfavorable. Because electron count and layer stacking change simultaneously, their individual roles remain unresolved, while electron-count tuning and alternative stacking arrangements offer possible routes to stabilize $AFe_6Ge_5$. The Fe–Mn stability preference is X-dependent, but the magnetic distinction among stable Ge compounds is clear: Fe-based members mainly favor AFM2 order, whereas Mn-based members consistently favor robust FM order. We identify six new stable FM candidates, making Mn-based Ge systems promising targets for synthesis. Their magnetic behavior is consistent with an interlayer exchange-competition mechanism analogous to that established in $AT_6X_6$ kagome magnets, while $ScFe_6Ga_4$ remains a boundary case that merits further noncollinear investigation. Representative FM compounds such as $ScMn_6Ge_4$ and $ScMn_6Ge_5$ retain kagome-derived band features near K, although the $AT_6X_5$ phase exhibits stronger band folding and hybridization. Overall, this work expands the kagome-magnet design space beyond $AT_6X_6$ and provides a practical framework for balancing thermodynamic stability, magnetic order, and kagome-derived electronic features.

**Acknowledgements**

The work at Xiamen University was supported by the National Natural Science Foundation of China (Grant No. T2422016). The work of V.A. was supported by the DOE Established Program to Stimulate Competitive Research (EPSCoR) Grant No. DE-SC0024284. Ames National Laboratory is operated for the U.S. Department of Energy by Iowa State University under Contract No. DE-AC02-07CH11358.